# Theoretical analysis of nanoparticle-induced homeotropic alignment in nematic liquid crystals

Amit Choudhary,[c†] Thomas F. George,[d] and Guoqiang Li [a,b,c*]

[a] *Department of Electrical and Computer Engineering,*

*The Ohio State University, 1330 Kinnear Road, Columbus, OH 43212, USA*

[b] *Department of Ophthalmology and Visual Science,*

*The Ohio State University, 1330 Kinnear Road, Columbus, OH 43212, USA*

[c] *College of Optometry, University of Missouri–St. Louis, St. Louis, MO 63121, USA*

[d] *Office of the Chancellor and Center for Nanoscience,*

*Departments of Chemistry & Biochemistry and Physics & Astronomy,*

*University of Missouri – St. Louis, St. Louis, MO 63121, USA.*

*\*Email: li.3090@osu.edu*

**Abstract:** A theoretical analysis of homeotropic alignment induced by nanoparticles (NPs) in a nematic liquid crystal (NLC) sample cell is presented. It is found that such alignment on the surface of a NP causes a change in the orientation of the molecular director near the surface, which in turn induces variations in the elastic constants and free energy. The induced NLC properties allow coupling between nearby NPs, mediated by the NLC molecules. The rotation of the coupled NPs close to the substrate tends to induce a long-range orientation of the NLC molecular director, leading to modification in the alignment at the interface of NLC and substrate which induces the orientation from homogeneous (planar) to homeotropic (vertical) in the bulk material.

## I. Introduction

The techniques to align a nematic liquid crystal (NLC) molecular director permit a wide range of applications of NLC materials other than displays. There are three types of basic alignments: homogeneous, homeotropic, and tilted. For homogeneous alignment, the molecules are parallel to the substrate of the sample cell. This has potential applications in switching of the NLC materials with positive anisotropy. In homeotropic alignment, the director is perpendicular to the substrate of the sample cell. The materials with negative anisotropy are preferred for this alignment so that the electric field parallel to the director can produce a Freedericksz transition of the NLC material. Tilted alignment is an intermediate state between homogeneous and homeotropic. The basic principle of alignment of the director inside the sample cell can roughly be explained by using the semiempirical Friedel-Creagh-Kmetz (FCK) rule,[1] which says that if the excess energy of a solid substrate is greater than the surface tension of NLC, then homogeneous alignment is preferred, and if it is the reverse, then homeotropic alignment is preferred. The most widely accepted technique to achieve uniform and thermally stable homogeneous alignment is polyimide film rubbing, which contributes to the orientation of NLC molecules parallel to the substrate.[1-4] Azo dye has also been adopted as one of the guiding materials to align the director optically,[5-13] along with other methods such as the application of a magnetic field.

Recently, nanoparticles (NPs) have also been utilized to achieve homeotropic alignment of the director.[14,15] The realization of NPs suspended in various host materials becomes essential because of their considerable influence on the core properties of the host LC materials, such as speeding up the switching of molecules, reducing the driving voltage of the device,[16,17] and generating a new mode.[18] The inclusion of NPs in NLC has been studied theoretically and experimentally in regard to alignment of NLC molecules around a NP. Studies in the past two decades have focused on several aspects such as topological defect formation (Saturn rings, hedgehogs, and boojums) in bulk NLC,[19-27] nematic-to-isotropic phase transitions,[28] self-assembly structure of NP mediated by NLC,[29] and control over the alignment of bulk LC induced by NP without treating the substrate surface. Polyhedral oligomeric silsesquioxanes (POSS)[14] and gold[15] nanoparticles with a few nanometers in size are an efficient tool to align the bulk NLC director. The surface capping agent, such as alkyl thiol derivatives, is used to protect the aggregation of NPs and tends to align the NLC molecules homeotropically on the surface of a NP, which becomes helpful in producing homeotropic alignment of the bulk NLC director.[30]

In POSS-doped NLC, spontaneous homeotropic alignment has been observed. The concentration of nanoparticles has been found to induce a variation in the pre-tilt angle, and after a concentration of ~1 wt%, the alignment becomes completely homeotropic.[31] The results have been interpreted that the POSS NPs could be absorbed on the substrate surface and reduce the surface energy of the substrate, subsequently producing the homeotropic alignment. Several studies, as mentioned above, have been carried out on the molecular alignment of NLC around NPs, but the focus has not been on the mechanism of homeotropic alignment of bulk NLC induced by the NPs within the confined geometry of the cell.

In this study, we carry out a theoretical analysis of the homeotropic alignment of bulk NLC material induced by NPs close to the substrate. We demonstrate the enhanced local ordering of NLC molecules on the surface of a NP due to the fact that homeotropic anchoring contributes largely to inducing homeotropic alignment of bulk NLC material within the confined geometry of the sample cell. The homeotropic anchoring of NLC molecules on the spherical surface of a NP shows the variation in elastic properties and elastic free energy, which are studied by assuming a contribution by smectic layer formation at the interface of nematic phase and solid surface.[32] The enhanced elastic properties have helped enable the coupling of two NPs in close proximity to act as a pair.[33] The NP pair tends to rotate the NLC alignment at the interface of NLC and substrate. The theory and analysis is presented in the next section, which includes the introduction of elastic constants, the elastic free energy density, and nanoparticle alignment and liquid crystal ordering. This is followed by the conclusions.

**II. Theory and analysis**

**A. Introduction of elastic constants**

For our study, the long and short range elastic properties of NLCs are described by three main elastic constants, $K_{11}$, $K_{22}$, and $K_{33}$, corresponding to splay, twist, and bend, respectively.[34] The elastic properties of NLC are analyzed in the context of NPs, so that considering the center of a NP at the origin, a single quadrant has been used for the analysis throughout the study, with the properties assumed as symmetric in the remaining quadrants. The bulk director is taken along the symmetry axis ($z$ axis), and the alignment of the NLC molecules on the surface of the NP are considered to be homeotropic.[19]

Consider the spherical coordinate system for the local director ($\vec{n}$) configuration with slight modifications, as shown in Fig. 1. The present system is axially symmetric, so that the director $\vec{n}$ depends on the polar angle $\theta$ and the angle $\beta$, which gives the local orientation of $\vec{n}$ with respect to the symmetry axis.[19] The molecular director components can be described as

$$n_r = \cos[\beta(r,\theta)]\,, \quad (1a)$$

$$n_\theta = \sin[\beta(r,\theta)]\cos\varphi\,, \quad (1b)$$

$$n_\varphi = \sin[\beta(r,\theta)] \sin\varphi, \tag{1c}$$

$$\vec{n} = n_r \hat{e}_r + n_\theta \hat{e}_\theta + n_\varphi \hat{e}_\varphi, \tag{1d}$$

$$\beta(r,\theta) = \theta - \frac{1}{2} \arctan\left[\frac{\sin(2\theta)}{(a/r)^3 + \cos(2\theta)}\right], \tag{1e}$$

where $n_r$, $n_\theta$, and $n_\phi$ are the unit vectors of the director at a point in the spherical coordinate system along $r$, $\theta$, and $\phi$, and $\hat{e}_r$, $\hat{e}_\theta$, and $\hat{e}_\phi$ are the local coordinate basis along $r$, $\theta$, and $\phi$, respectively. The parameter $a$ is the radius of disclination singularity around NP in NLC,[19] where the disclination ring is the ring-shaped structure formed by the singularity of molecular director around the NP. The term singularity means that it is a region where the magnitude of nematic order goes to zero. It has been shown previously[19,35] that the deviation angle of the director of the NLC molecules at an arbitrary position depends on position vector $r$, and its polar angle $\theta$ with respect to the main director of the bulk NLC, as reflected in Eq. (1e). The angle $\beta$ is almost equal to the polar angle $\theta$ close to the periphery of the NP surface. This is due to the strong homeotropic coupling of the long axis of the NLC molecule on the surface of the NP, but away from the NP, the angle is directed by the director of the bulk NLC material.

In general, the elastic constants ($K_{11}$, $K_{22}$, and $K_{33}$) can be defined by the method given in Refs. [36, 37]. Due to the complexity in applying them for practical utilization, another method has been proposed in Ref. 38. The behavior of these constants are defined for an undoped bulk NLC material, where the smectic-like phase formation at the interface of solid substrate as shown in Fig. 2 is not considered. The smectic-like phase formation by nematic molecules at the interface of an isotropic solid surface with NLC had been proposed earlier by de Gennes.[32] This factor needs to be considered in explaining the variation in the twist and bend elastic constants of NLC in the vicinity of a second-order nematic-to-smectic A phase transition. Therefore, new terms $K'_{22}$ and $K'_{33}$ have been defined in order to reformulate the elastic constants $K_{22}$ and $K_{33}$:[32, 39]

$$K'_{22} = \frac{\pi k_B T \xi_\perp^2}{6 d_o^2 \xi_\parallel}, \tag{2}$$

where $\xi_\perp = \frac{6}{\left\{\frac{(T-T_c)}{T}\right\}^{1/2}}$, $\xi_\parallel = 4\xi_\perp$, $d_o \approx 2$ nm, and

$$K'_{33} = \frac{1}{2} \frac{\pi k_B T}{6 d_o^2} \xi_\parallel. \tag{3}$$

Here $\xi_\perp$ and $\xi_\parallel$ are the coherent lengths of the smectic phase formation perpendicular and parallel, respectively, to the layer of NLC formation at the interface of the solid surface and nematic, and $T$ and $T_c$ are the regular temperature and transition temperature of the sample. These parameters are dependent on the temperature of the NLC phase. The layer thickness, $d_o$, is the projection of the average molecular length in the smectic layer formed at the interface of the NLC and solid substrate due to the strong molecular anchoring at the interface. (Note: The schematic in Fig. 1(a, b) shows the direction of the $z$ axis to represent the normal for any kind of solid surface, but the initial bulk director throughout the rest of the paper will be considered along the substrate surface, i.e., homogeneous alignment.)

The expressions for $K'_{22}$ and $K'_{33}$ in Eqs. (2) and (3) are applicable to flat substrate surfaces. They may not be appropriate for other types of surfaces such as cylindrical and spherical. In the case of a NP, the surface is assumed to be spherical. The homeotropic attachment of NLC molecules is also in spherical manner constituting the topological defect.[19] Therefore, $K'_{22}$ and $K'_{33}$ must be further modified, and in order to do this, it is necessary to analyze the NLC molecular alignment at the periphery of the NP. The orientation ordering of NLC molecules close to the surface of the NP can vary as a function of $r$ and $\theta$ as shown in Fig. 1. Then, the modified terms will be a function of these two parameters. We can refer to these modified constants as elastic constants for curved surfaces, i.e., $K_{22}^{curve}$ and $K_{33}^{curve}$. A function $f(r,\beta)$ needs to be introduced into $K'_{22}$ and $K'_{33}$ as

$$K_{22}^{curve} = \frac{\pi k_B T \xi_\perp^2}{6 d^2 \xi_\parallel} f(r,\beta), \tag{4}$$

$$K_{33}^{curve} = \frac{1}{2} \frac{\pi k_B T}{6 d^2} \xi_\parallel f(r,\beta), \tag{5}$$

where $f(r,\beta) = \left(\frac{R}{r}\right)^2 \cos^2\beta$, $d \approx d_o + \frac{(r-R)^2}{d_o} \cos^2\theta$, and $T$ is the temperature of the NLC. The smectic layer formation at the interface loses its existence as a function of $r$ and $\theta$ because of NLC molecular alignment and its fluidic nature. Therefore, the smectic-like layer spacing is a variable quantity as $d \sim d_o + (r - R)^2 \cos^2\beta/d_o$ rather than a constant one, where $R$ is the radius of the NP. Then, the modified expressions for the elastic constants of the systems can be written as

$$K_{11} = \left[1 + \lambda_1 - 9\lambda_1 z_1 + \left(6 + \frac{39}{11}\lambda_1\right) z_1^2\right] \times K, \tag{6a}$$

$$K_{22} = \left[1 - 2\lambda_1 - 3\lambda_1 z_1 + \left(6 - \frac{141}{11}\lambda_1\right) z_1^2 + K_{22}^{curve}\right] \times K, \tag{6b}$$

$$K_{33} = \left[1 + \lambda_1 + 12\lambda_1 z_1 + \left(6 + \frac{102}{11}\lambda_1\right) z_1^2 + K_{33}^{curve}\right] \times K, \tag{6c}$$

where $\lambda_1 = \frac{2}{7}\omega$, $\omega = \frac{\gamma^2 - 1}{\gamma^2 + 1}$, $\gamma = \frac{\sigma_\parallel}{\sigma_\perp}$, $z_1 \approx \frac{\bar{P}_4}{\bar{P}_2} \approx \frac{1}{4}\left[\frac{35 \cos\beta - 30 \cos^2\beta + 3}{3\cos^2\beta - 1}\right]$, and $K = \frac{1}{3}(K_{11} + K_{22} + K_{33})$, where $K \sim 10^{-11}$ Newton is an approximate value for the arithmetic mean of the three elastic constants of NLC.

It is well established that the homeotropic alignment of NLC molecules around the NP surface constitutes a topological defect and a disclination ring of certain radius '$a$' greater than the radius of NP.[19] In the region of disclination, the function of the elastic constant becomes undefined and shows the low and high magnitudes of the respective properties, as illustrated in Fig. 3, where the radial distance $r^*$ in reduced unit is used to make the simulation simpler. The dependence of the elastic constant $K_{22}/K$ on the disclination ring in Fig. 3 shows that at 45° it decreases and then tends to the value equivalent to that at 0° and 90°. In Fig. 3(d), the maximum value of $K_{22}$ is $3.84 \times 10^{-11}$ N when $\theta = 0°$ and $r^* = 22.2$. However, the behavior of $K_{33}/K$ in Figs. 4(a-c) shows a strong dependence at 45° as a function of the position vector $r$. The maximum value $1.23 \times 10^{-11}$ N of $K_{33}$ at 0° and 22.2 is obtained as shown Fig. 4(d), where $K_{33}/K$ first decreases and then increases due to the asymptotic nature of the singularity. The peaks in Figs. 3 and 4 at the position of the disclination radius are due to the asymptotic behavior of the singularity at the distance equal to the ring radius. The singularity is the region where the function becomes undefined. The shift of the disclination ring as a function of $r$ and $\theta$ is dependent on the strength of the anchoring energy of the NP surface, which holds the NLC molecules aligned homeotropically at its surface up to a certain value of the distance $r$ and angle $\beta$. The alignment of NLC molecules between the surface of the NP and the disclination ring remains homeotropic with respect to the NP surface. The molecules tend to remain homeotropic outside the ring to some extent at $\theta \sim 90°$ due to the elastic effect. Consider a large disclination ring such that we then have a large defect due to the strong homeotropic alignment of NLC molecules on the surface of the NP. Figures 3(d) and 4(d) show that the strong stress is generated at a position close to the particle surface with the molecules parallel to the bulk director. This is due to the enhanced values of $K_{22}$ and $K_{33}$ which are modified by introducing $K_{22}^{curve}$ and $K_{33}^{curve}$, respectively.

## B. Elastic free energy density

The generated elastic forces in the NP regime can be visualized through the elastic free energy as well, as shown in Fig. 5. Elastic free energy is that amount of energy in the system which is available for work at the equilibrium state due to distortion in ordering. For an analysis of this, we consider the molecules to be anchored homeotropically on the surface of the NP, generating a Saturn ring singularity pattern of the radius director *a* perpendicular to the bulk alignment. For estimation of the elastic free energy using Eqs. (1) and (6), we select one quadrant of the spherical coordinate system, with the other quadrants of the NP sphere remaining symmetric. The elastic free energy configuration of NLC can be defined as[34, 35]

$$f_e = \frac{1}{2}K_{11}|\nabla \cdot \vec{n}|^2 + \frac{1}{2}K_{22}|\vec{n} \cdot \nabla \times \vec{n}|^2 + \frac{1}{2}K_{33}|\vec{n} \times \nabla \times \vec{n}|^2. \tag{7}$$

As shown in Fig. 5, the free energy density reaches the maximum value of $4.66 \times 10^{-9}$ J/m$^2$ when $\theta = 0°$ and $r^* = 22.2$. The large value for $\theta$ between 0.8 and 1.5 (rad) is due to disordered alignment (i.e., disclination). It is well known that the director in the bulk NLC is aligned according to the substrate surface alignment restriction, but in the close proximity to the NP, it is almost perpendicular to the surface. In general, the director is oriented either by surface effects or by an external electric or magnetic field. The effects of the substrate surface and external field are not considered in the elastic free energy expression because the assumed systems under study are NP-doped and undoped NLC samples.

The angle $\beta$ between the local director *n* close to the NP and the director of the bulk NLC, i.e., along the *z* axis in the present case (Fig. 1), generates the variation in the elastic free energy close to the NP, which is dependent on the polar angle $\theta$ and position vector *r* from the center of the NP, according to Eq. (7). The value of $\beta$ is almost equal to $\theta$ near the periphery of the NP due to the strong homeotropic coupling of the long molecular axis of the NLC molecules on the surface of the NP; but away from the NP, it is guided by the alignment of the bulk director. The variation of the elastic constants as a function of $\beta$, shown in Figs. 3(d) and 4(d), represents the deviation in intermolecular interactions and produces the subsequent change in the elastic free energy. Figure 5 shows the deviation in the free energy at the periphery of the NP. The elastic free energy in the region close to $r \sim R$ (nm) and $\beta \sim \theta$ (radian) is higher in comparison to the bulk elastic free energy away from the NP due to the strong anchoring of the NLC molecules. In this region, the NLC molecules have higher interaction and are aligned forcibly by the anchoring effect of the NP surface. Any other particle (NP or surface of the substrate of the sample cell, which is the special case of a particle having infinite radius), at this periphery of high interaction energy will be dealt with according to the deviated energy profile. These interactions of NLC molecules with other particles in this energy profile area are assumed to be responsible for the change in the properties of NLC materials. These particles are able to form pairs of NPs mediated by NLC molecules, while away from the NP they can behave independently.

**C. Nanoparticle alignment and liquid crystal ordering**

Two or more particles in the NP doped NLC materials interact with each other according to the following relation of particle energy[33]

$$U\left(|\vec{\ell}|,, \Psi\right) = \frac{\pi W^2 R^8}{30 K |\vec{l}|^5}\left(1 - \frac{WR}{56K}\right)[9 - 20\cos(2\Psi) + 35\cos(4\Psi)], \qquad (8a)$$

where $W$, $R$, $K$, $|\vec{l}|$, and $\Psi$ are the anchoring energy of the NP surface, NP radius, average elastic constant under one constant approximation (i.e., an average value of $K$ has been considered for the whole NLC material system), spacing between the two NPs (since this distance is orientation dependent, a vector notation is given), and the angle formed by the line between the NPs with respect to the direction of the bulk alignment, respectively. For particles that are small enough, $WR/56K \ll 1$. Then Eq. (8a) reduces to

$$U\left(|\vec{l}|,, \Psi\right) \approx \frac{W^2 R^8}{K |\vec{l}|^5}[0.914 - 2.094 \cos(2\Psi) + 3.665 \cos(4\Psi)]. \qquad (8b)$$

It has been assumed that initially the two particles come closer due to gravity and interact with each other effectively when they are under the higher interaction energy profile area. The following values have been used to analyze the pair interaction function as shown in Fig. 6:

$W \cong 10^{-6}$ J/m², $|\vec{l}| = 5 - 50$ nm, $\Psi = 0 - \frac{\pi}{2}$ (radian), and $K \sim 10^{-11}$ N.

Figure 7 shows that the NPs close to each other possess a conditional interaction so that the interaction parallel and perpendicular to the molecular director is repulsive, whereas at other angles, it is attractive. We assume an NP to experience a gravitational potential and reach equilibrium near the substrate surface due to buoyancy force exerted by NLC on the NP. The equilibrium height, $h$, of the NP from the substrate of the sample cell depends on the elastic strength of the NLC material. If the NLC molecules are strongly anchored to the surface of the NP, then it is strongly repelled upward due to the elastic energy of the NLC material, $U \sim 2\pi^2 K R^4/h^3$.[40] If they were not repelled, then after sufficient time there is a possibility to achieve equilibrium through the gravitational potential energy, $U_g \sim (4/3)\pi R^3 \Delta\rho g h$, where $\Delta\rho = \rho_p - \rho_N$, $\rho_p$ and $\rho_N$ are the density of the particle and the NLC material, respectively, and g is the acceleration due to gravity. Thus, this force is balanced by the buoyancy of NLC on the NP at the equilibrium height, $h$.[40]

Considering the bulk alignment of the NLC to be homogeneous within the sample cell, one can obtain an expression for the effective height of the NP from the sample substrate. Initially, the angle between the NLC director and the substrate surface is assumed to be *0º*, i.e. homogeneous alignment, as shown in Fig. 7. When the angle $\Psi$ between the distance line $\overrightarrow{|l|}$ of the two particles (one is NP and the other is the substrate itself) and the bulk alignment is ~90º, the pair potential gives a repulsive force, and at ~45º, it gives an attractive force.

Since the initial alignment of the bulk director is considered homogeneous, the assumed angles also correspond to the homogeneous alignment. The balancing energy equation can be written as

$$U_g + U_{attractive} = U_{repulsive}, \qquad (9)$$

where $U_g$, $U_{attractive}$, and $U_{repulsive}$ are the gravitational, attractive, and repulsive pair potential energies, respectively. By substituting the expressions for the gravitational potential,[40] attractive and repulsive energies from Eq. (8b) and Eq. (9), the equilibrium height can be found as

$$h = \left[\frac{W^2 R^5}{20 \Delta\rho g K}\left(1 - \frac{WR}{56K}\right)\right]^{1/6}. \qquad (10)$$

The stronger the anchoring of the NLC on the NP and the smaller the NP radius, the higher is the hanging position of the NP on the NLC close to substrate. If $h$ is less than the disclination radius, *a,* the molecular director between the substrate and the NP tends to be homeotropic. This will not allow the NLC molecules to wet the substrate surface because of the strong inward pull by the NP surface interactions. Such a pull creates a depletion-type region between two particles,[41] where the combination forces tend to push the molecules aside. In the present case, one end of the NLC molecule is strongly coupled to the NP and the other end is loosely bound to the substrate. This combination of anchoring of the NLC molecules contributes to the conversion of the molecular director from the homogeneous to the homeotropic state. This results in the change of surface tension because of the inward pull of NLC molecules by the NP.

Assuming homogeneous alignment of the bulk NLC, one NP (called NP-1) is in equilibrium at some height $h$ due to repulsive and attractive interactions, as shown in Fig. 8. Another NP (NP-2) comes into the contact with NP-1 under the attractive and repulsive pair potentials at various angles and the gravitational potential due to the substrate. The two NPs experience an orientation mediated by the elastic NLC material, and there is a strong anchoring of the NLC at its surface.

This orientation of NPs tends to rotate the NLC director associated with it in the effective region of higher energy as per Eq. (7) and is up to longer distances because of the elastic and anisotropic properties of the NLC material. The pair formed tends to acquire a stable state according to the oriented attractive and repulsive potentials of NP-1 and NP-2. Then, the torque applied on NP-1 by NP-2 under the gravitational motion can be expressed as

$$\vec{\Gamma}_{12} = \vec{l} \times (\vec{F}_g + \vec{F}_{pair} - \vec{F}_{substrate}), \qquad (11)$$

where $F_g$, $F_{pair}$, $F_{substrate}$, and $|\vec{l}|$ are the forces generated by the gravitational effect, the NP pair formation, the substrate to produce the torque on the NP-1 by NP-2, and the distance between the two NPs in the form of the vector, respectively. Substituting the expression for the force from the calculation of respective potentials, Eq. (11) can be written as

$$\vec{\Gamma}_{12} = \frac{4}{3}\pi R^3 \Delta \rho g C |\vec{l}| \sin\Psi + \frac{109\pi R^8}{9\sqrt{2}K|\vec{l}|^5}\left(1 - \frac{WR}{56K}\right). \qquad (12)$$

We have introduced the term $C$ as the ratio of forces applied by the NLC molecules on each other under the elastic deformation and the force applied on the NLC molecules under the influence of the uniform bulk molecular director:

$$C = \frac{K_{11}|\nabla \cdot \vec{n}| + K_{33}|\vec{n} \times \nabla \times \vec{n}|}{3\mu_i \mu_j/(2\pi \varepsilon r_{ij}^2)}. \qquad (13)$$

Here, $\mu_i$ and $\mu_j$ are the dipole moments of the $i^{th}$ and $j^{th}$ molecules in the bulk LC, respectively, $r_{ij}$ is the relative distance between them, and $\varepsilon$ is the dielectric constant of the NLC material. The specific elastic constants, $K$'s, can be considered in the calculation of $C$ instead of taking its average value to show the anisotropic effect of molecular anchoring around the NPs. The torque on the two NPs, mediated by the ordering of the NLC molecules, also rotates the NLC molecules to reach equilibrium in the NLC system. Thus, the two particles NP-1 and NP-2 are now in equilibrium close to the substrate surface, as shown in Fig. 8(a). The substrate of the sample cell also acts as a particle. Now suppose a third particle NP-3 comes between NP-1 and NP-2, as shown in Fig. 8(b). It will also be in equilibrium with the fourth one, i.e. the substrate which is also acting as a particle of very large surface area. Before achieving equilibrium, NP-3 induces a slight orientation of the NP-1 and NP-2 pair. Then NP-1, NP-2 and NP-3 will experience the same interaction with the substrate that they could experience from any other particles.

The interaction of the NLC with the substrate can be defined by the potential expression[42]

$$\Phi(u_i, x_i) = \frac{2\,\varepsilon_s}{3\sigma_s^2}\left[\frac{2}{15}\left(\frac{\sigma_o}{\sigma_s x_i}\right)^9 - \left(\frac{\sigma_o}{\sigma_s x_i}\right)^3\right], \tag{14}$$

where $\varepsilon_s$ is the well depth strength parameter and depends on $\gamma$, the intermolecular size parameter and is also another parameter which can be used to adjust the ratio between the end-end and side-side potential well depth. Let us consider the special case in which $x_i$ is the distance of the $i^{th}$ molecule from the substrate, and $u_{i,z}$ is the orientation unit vector of the $i^{th}$ molecule with $z$ axis. Then, the parameters in Eq. (14) can be written as

$$\sigma_s = \left[1 - \frac{\chi u_{i,x}^2}{1-\chi^2 u_{i,z}^2}\right]^{-1/2}, \tag{15}$$

$$\chi^2 = \left(\frac{\gamma^2-1}{\gamma^2+1}\right), \tag{16}$$

$$\gamma = \frac{\sigma_\parallel}{\sigma_\perp}, \tag{17}$$

$$\varepsilon_{os} = \varepsilon\varepsilon_o, \tag{18}$$

$$\varepsilon_s = \varepsilon_{os}[1 - \chi^2 u_{i,z}^2]^{-1/2}, \tag{19}$$

where $u_{i,x}$ is the orientation of the $i^{th}$ molecule along the $x$ axis, $\Phi_{NP}$ and $\Phi_s$ are the potentials of the NP and the substrate, respectively, and can be obtained by Eq. (14). If $\Phi_s < \Phi_{NP}$, then the NLC molecules on the substrate can be easily rotated by the NP pair orientation near the substrate surface. Assume $F_{NLC\text{-}substrate} = -\nabla \Phi_s (u_i, x_i)$ and $F_{NLC\text{-}NP} = -\nabla \Phi_{NP} (u_i, x_i)$ are the forces exerted by the substrate and the NP on the NLC molecules at equilibrium, respectively. If the condition $F_{NLC\text{-}NP} > F_{NLC\text{-}wall}$ is satisfied, the final molecular torque produced by the pair orientation, elastic effect, and wall interaction can effectively rotate the NLC molecules in the orientation direction of NP pair, as shown in Fig. 8. The orientation of the NLC molecules in between the NPs, shown in Fig.8 (b), creates a long-range interaction and high stress inside the NLC material. This stress results in an enhancement in the surface tension of the NLC material. The molecules of such a NLC system do not experience enough anchoring on the substrate surface, leading to adherence to the FCK rule,[1] i.e., homeotropic alignment on the substrate surface. The conversion of alignment depends on the concentration of the NPs in the NLC material. The pre-tilt dependence on the concentration of NPs in the NLC has been shown by Hwang *et al*.[43]

The behavior of the tilt angle in the doped NLC samples can be understood by considering the torque balance equation of the oriented NLC molecules,[34]

$$\vec{\Gamma}_{elastic} + \vec{\Gamma}_{pair} - \vec{\Gamma}_{substrate} = 0, \tag{20}$$

where $\Gamma_{elastic}$, $\Gamma_{pair}$, and $\Gamma_{substrate}$ are the contributions of torque to rotate the molecule from the elastic properties of NLC, pair of NPs, and substrate, respectively. The torque by the substrate, $\Gamma_{wall}$, can be considered negligible because of its static nature, so that

$$K_{22}\frac{\partial^2 \phi}{\partial z^2} + \nabla U \times \vec{n} = 0, \tag{21}$$

where $\phi$ is the tilt angle from the lower to upper substrate. The potential, $U$, produced by the pair acts as a source of the force to generate the torque on the unit vector $\vec{n}$ of the NLC molecule in between the two particles of a pair. $\Gamma_{substrate}$ is negligible at large distances from the substrate. So, the specific equations can be obtained by using various values of $\Psi$ in Eq. (8).

If the angle between the distance line $\vec{|l|}$ of the two NPs with respect to the bulk director is $\Psi \sim 45°$ in Eq. (8b), then Eq. (21) is reduced to

$$K_{22}\frac{d^2\phi}{dz^2} + \frac{13\pi}{3}\frac{W^2 R^8}{K_{22}|\vec{l}|^6}\left(1 - \frac{WR}{56K_2}\right)\sin\phi = 0, \tag{22}$$

and if $\Psi \sim 90°$ in Eq. (8b), the potential energy is repulsive and the torque balancing equation reduces to

$$K_{22}\frac{d^2\phi}{dz^2} - \frac{13\pi}{3}\frac{W^2 R^8}{K_{22}|\vec{l}|^6}\left(1 - \frac{WR}{56K_2}\right)\sin\phi = 0. \tag{23}$$

Using boundary conditions at $z = 0$ and $t$ (thickness of the sample cell), the magnitude of $\phi$ will be around 0 and 90° for homeotropic and planer alignment, respectively. At $\Psi \sim 90°$, the particles will be in equilibrium near the substrate. The solutions for $\phi$ in Eqs. (22) and (23) define the path of the molecular director orientation from $z = 0$ to $t$ as a function of the elastic constant and the anchoring energy of NLC molecules on the NP surface.

## III. Conclusions

In summary, we have shed light on NP-induced homeotropic alignment of a NLC material. The smectic-like layer formation in close proximity to the NP leads to variation in the elastic properties and free energy of the NLC material. Modulation in the properties of the NLC allows the pair formation of NPs close to the substrate surface, which is mediated by the NLC molecules. Such pair formation results in the conversion of alignment of NLC molecules between NLC and the substrate. Thus, the surface tension increases due to the conversion of molecular alignment close to the surface of the substrate. This results in adherence to the FCK rule of alignment conversion from homogeneous to homeotropic.

## AUTHOR INFORMATION

**Present addresses**

[†] Present Addresses: Physics Department, Deshbandhu College, University of Delhi, Kalkaji, New Delhi-110019, India.

**Acknowledgments**

This work was supported in part by the National Institutes of Health National Eye Institute, USA (through grant R01 EY020641), National Institute of Biomedical Imaging and Bioengineering, USA (through grant R21 EB008857), National Institute of General Medical Sciences, USA (through grant R21 RR026254/R21GM103439), and Wallace H. Coulter Foundation Career Award (through grant WCF0086TN).

1. J. Cognard, Mol. Cryst. Liq. Cryst. (Suppl. Ser.) A5, 1 (1982).

2. *J. S. Patel, T.M. Leslie, and J. W. Goodby, Ferroelectrics* **59**, *137 (1984).*

3. *J. M. Geary, J. W Goodby, A. R. Kmetz, and J. S. Patel, J. Appl. Phys.* **62**, *4100 (1987).*

4. *D. W. Berreman, Mol. Cryst. Liq. Cryst.* **23**, *215 (1973).*

5. *W. M. Gibbons,* P. J. Shannon, S. T. Sun, and B. J. Swetlin, *Nature* **351**, *49 (1991).*

6. *K. Ichimura,* Y. Hayashi, H. Akiyama, T. Ikeda, and N. Ishizuki, *Appl. Phys. Lett.* **63**, *449 (1993).*

7. *S. T. Sun,* W. M. Gibbons, and P. J. Shannon, *Liq. Cryst.* **12**, *869 (1992).*

8. M. Eich, J. H. Wendorff, B. Reck, and H. Ringsdorf, Makromol. *Chem. Rapid Commun.* **8**, *59 (1987).*

9. *M. Schadt,* H. Seiberle, A. Schuster and S. *M*. Kelly, Jpn. J. Appl. Phys. **34**, L764 (1995).

10. M. O'Neill and S. M. Kelly, J. Phys. D: Appl. Phys. **33,** R67 (2000).

11. K. Ichimura, Y. Suzuki, T. Seki, A. Hosoki, and K. Aoki, Langmuir **4**, 1214 (1988).

12. K. Ichimura, Y. Suzuki, T. Seki, Y. Kawanishi, T. Tamaki, and K. Aoki, Jpn. J. Appl. Phys. **28**, Suppl. 28-3, 289 (1989).

13. Y. Kawanishi, T. Tamaki, T. Seki, M. Sakuragi, and K. Ichimura, Mol. Cryst. Liq. Cryst. **218**, 153 (1992).

14. S. J. Hwang, S. C. Jeng, C. Y. Yang, C. W. Kuo, C. C.Liao, J. Phys. D: Appl. Phys. **42,** 025102 (2009).

15. H. Qi and T. Hegmann, Appl. Mater Interfaces **1**, 1731 (2009).

16. W. Lee, C. Y. Wang and Y. C. Shih, Appl. Phys. Lett. **85,** 513 (2004).

17. C. Y. Huang, H. C. Pan, and C. T. Hsieh, Jpn. J. Appl. Phys. **45,** 6392 (2006).

18. Y. Shiraishi, N. Toshima, K. Maeda, H. Yoshikawa, J. Xu, and S. Kobayashi, Appl. Phys. Lett. **81,** 2845 (2002).

19. O. V. Kuksenok, R. W. Ruhwandl, S. V. Shiyanovskii, and M. Terentjev, Phys. Rev. E, **54**, 5198 (1996).

20. J. Fukuda, M. Yoneya, and H. Yokoyama, Phys. Rev. E **65**, 041709 (2002).

21. Y. Gu and N. L. Abbott, Phys. Rev. Lett. **85**, 4719 (2000).

22. P. Poulin, H. Stark, T. C. Lubensky and D. A. Weitz, Science **275**, 1770 (1997); Phys. Rev. Lett. **79**, 4862 (1997); Phys. Rev. E **57**, 626 (1999).

23. R. Yamamoto, Phys. Rev. Lett **87**, 075502 (2001).

24. J. Fukuda, M. Yoneya, H. Yokoyama, Eur. Phys. J. E, **13**, 87 (2004).

25. M. A. Bates, Liq. Cryst. **32**, 1525 (2005).

26. I.I. Smalyukh, O. D. Lavrentovich, A. N. Kuzmin, A. V. Kachynski, and P. N. Prasad, Phys. Rev. Lett. **95**, 157801 (2005).

27. I. Musevic, M. Skarabot, U. Tkalec, M. Ravnik, and S. Zumer, Science **313**, 954 (2006).

28. S. K. Prasad, K. L. Sandhya, G. G. Nair, U. S. Hiremath, C. V. Yelamaggad, and S. Sampath, Liq. Cryst. **33**, 1121 (2006).

29. T. Hirai, M. Leolukman, C. C. Liu, E. Han, Yun J. Kim, Y. Ishida, T. Hayakawa, M. Kakimoto, P. F. Nealey, and P. Gopalan, Adv. Mater. **21**, 4334 (2009).

30. H. Qi, B. Kinkead, V. M. Marx, H. R. Zhang, and T. Hegmann, Chem. Phys. Chem. **10**, 1211 (2009).

31. S. C. Jeng, S. J. Hwang, and C. Y. Yang, Opt. Lett. **34,** 455 (2009).

32. P. G. de Gennes, Solid State Commun. **10**, 753 (1972); Mol. Cryst. Liq. Cryst. **21,** 49 (1973).

33. R. W. Ruhwandl and E. M. Terentjev, Phys. Rev. E **55,** 2958 (1997).

34. I. C. Khoo, *Liquid Crystals*, 2$^{nd}$ Ed. (Wiley Interscience, 2007).

35. H. Stark, Eur. Phys. J. B **10**, 311 (1999).

36. A. Poniewierski, and J. Stecki, Mol Phys. **38**, 1931 (1979).

37. J. G. Gay and B. J. Berne, J. Chem. Phys. **74**, 3316 (1981).

38. A.V. Zakharov and A. Maliniak, Eur. Phys. J. E **4**, 85 (2001).

39. K.C. Chu and W. L. McMillan, Phys. Rev A **11**, 1059 (1975).

40. E. M. Terentjev, Phys. Rev. E **51**, 1330 (1995).

41. R. Roth, R. van Roij, D. Ankrienko, K. R. Mecke, and S. Dietrich, arXiv:cond-mat/0202443x1[cond-mat.soft] 25 Feb 2002.

42. A. L. Tsykalo and A. D. Bagmet, Acta Phys. Pol. A **55**, 111 (1979).

43. S. J. Hwang, S. C. Jeng, and I. M. Hsieh, Opt. Exp. **18**, 16507 (2010).

**Figure captions:**

Fig. 1. Schematic representation of NP-doped NLC shown in a spherical coordinate system. One quadrant is considered in all the calculations.

Fig. 2. (a) Representation of smectic-like phase formation at the interface of a solid substrate and nematic phase, where $d_o$ represents the layer spacing close to solid NLC interface. (b) Smectic-like phase formation in NLC at the interface with a solid spherical NP surface.

Fig. 3. $K_{22}/K$ plotted as a function of the radial distance $r^*$ (in reduced units, $r^* = r/\sigma$, $\sigma = 0.450$ nm) from the center of the nanoparticle at different radii of the Saturn ring, $a$: (a) 33.3, (b) 73.3, and (c) 93.3 (in reduced units) and various polar angles ($\theta = 0°, 45°,$ and $90°$) for the singularity formed by the molecular director (Eq. (6b)). The behavior of $K_{22}/K$ and $K_{33}/K$ as a function of $r^*$ shows that it is clearly affected by the disordering of the NLC near the NP surface and the disclination defect. (d) $K_{22}$ plotted as a function of $r^*$ and $\theta$ in bulk NLC in the vicinity of NP (Eq. (6b)).

Fig. 4. $K_{33}/K$ as a function of the radial distance $r^*$ (in reduced units, $r^* = r/\sigma$, $\sigma = 0.450$ nm) from the center of the nanoparticle at different radii of the Saturn ring, $a$: (a) 33.3, (b) 73.3, and (c) 93.3 (in reduced units) and various polar angles ($\theta = 0°, 45°,$ and $90°$) for the singularity (Eq. (6c)) formed by the molecular director. The elastic constants have high values at first and then decrease near the disclination, becomings normal according to the bulk alignment. (d) $K_{33}$ plotted as a function of $r^*$ and $\theta$ in bulk NLC in the vicinity of the NP (Eq. (6c)).

Fig. 5. Elastic free energy plotted as a function of $r^*$ and $\theta$. The high jump peak values are due to the singularity structure in the Saturn ring. The azimuthal angle is kept constant at $\phi \sim 90^o$, and the radius of Saturn ring is approximately 73.3 (in reduced units).

Fig. 6. Schematic diagram of two NPs interacting via liquid crystalline matrix distortions. The arrow shows the direction of the force exerted by an NP on another nearby particle.

Fig. 7 Pair potential, $U(|\vec{l}|, \Psi)$, plotted as a function of the distance, $|\vec{l}|$, between two NPs and the angle, $\Psi$, formed by $|\vec{l}|$ with the direction of the bulk director (i.e., $z$ axis as shown in Fig. 6).

Fig. 8. (a) NP-1 settled at the equilibrium height $h$, and NP-2 falling under the gravitational effect from height $h'$, changing the orientation of NP-1 and itself to reach equilibrium, (b) NP-3 falling in between two already stable NPs, forming a zig-zag pattern through alternatively opposite interactions between NPs due to repulsion from other particles.

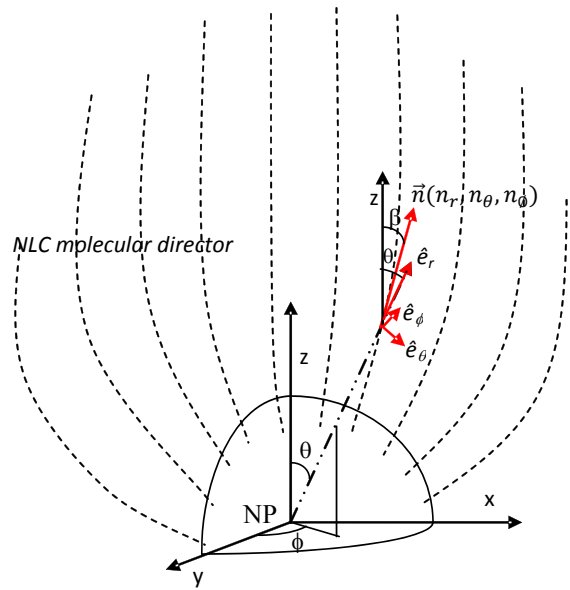

Fig. 1

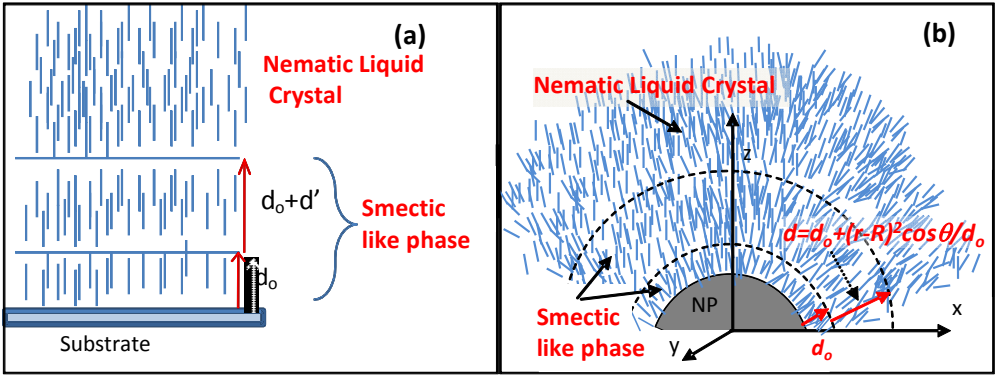

Fig. 2

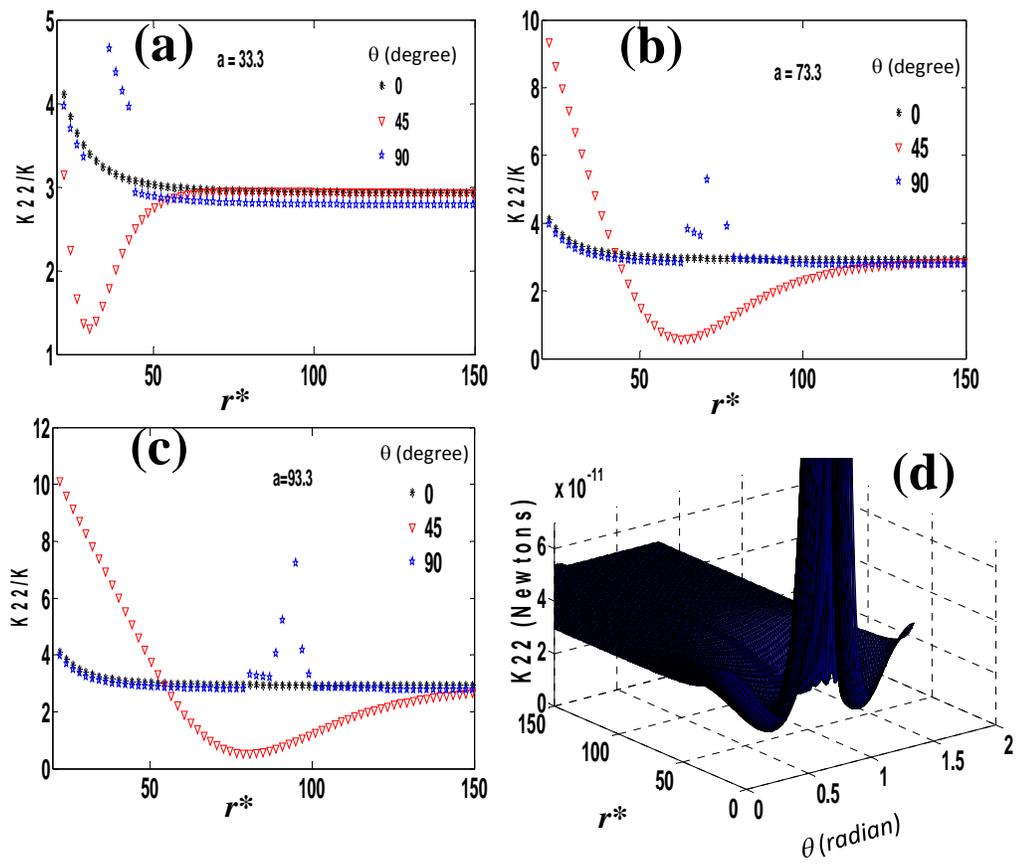

Fig. 3

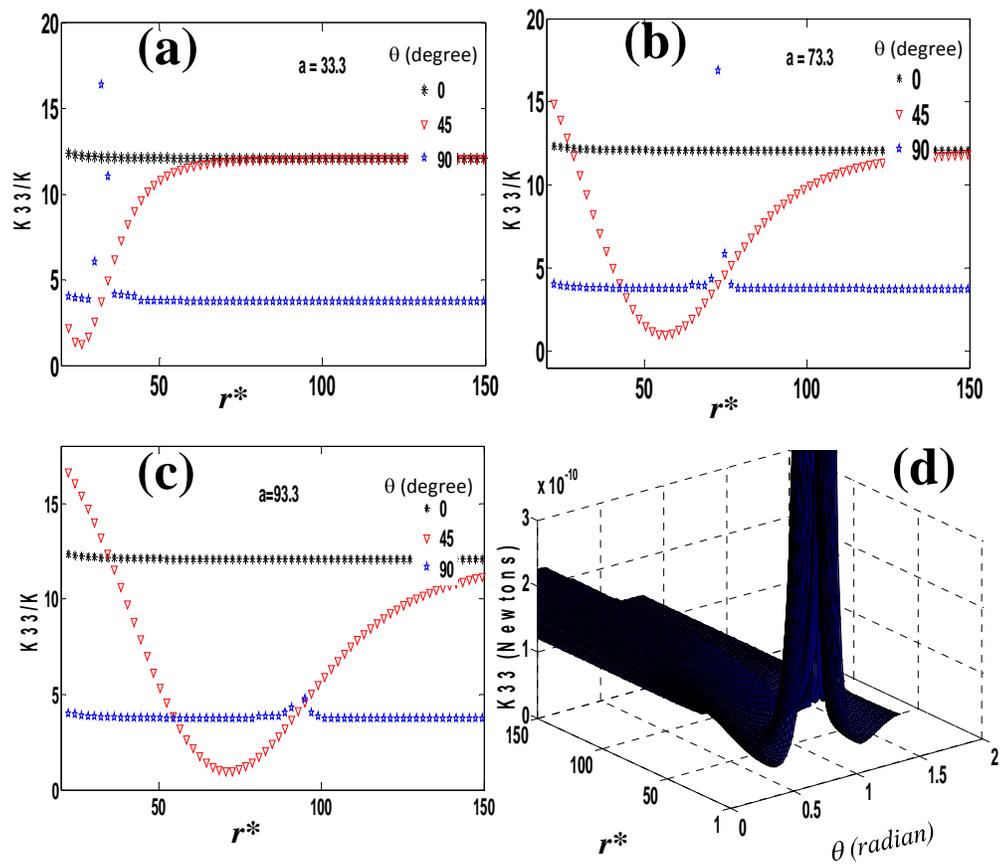

Fig. 4

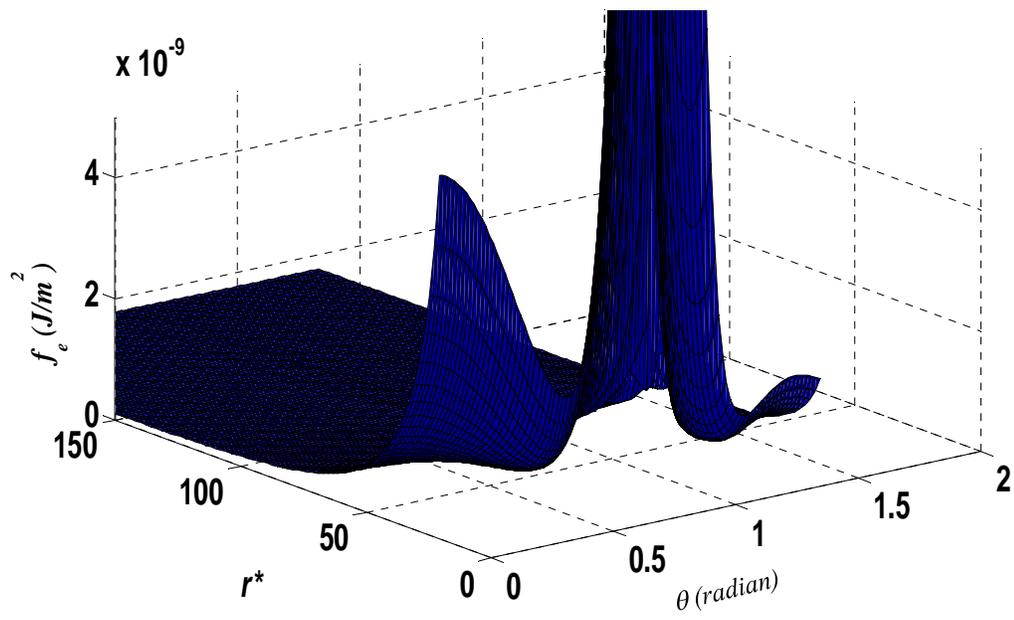

Fig. 5

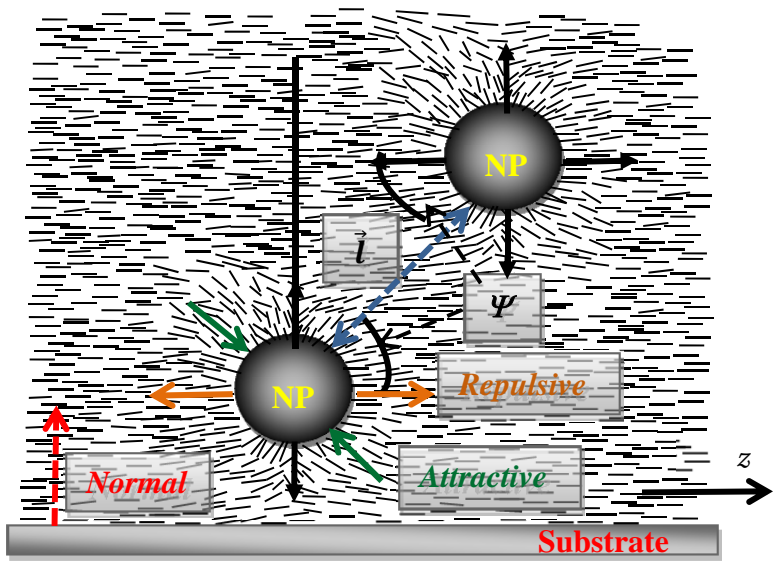

Fig. 6

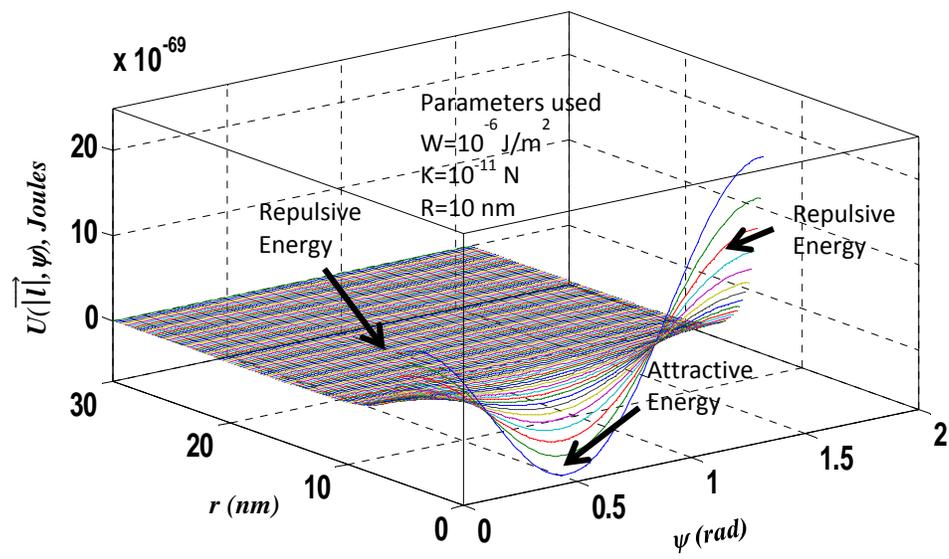

Fig. 7

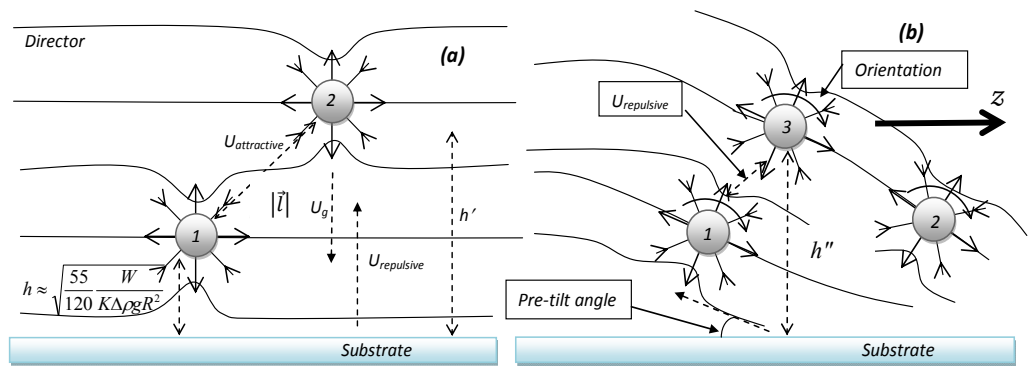

Fig. 8